\documentclass[10pt,draftcls,onecolumn]{IEEEtran}
\usepackage{amsmath}
\usepackage{amssymb}
\usepackage{upgreek}
\usepackage{tipa}
\usepackage{enumerate}
\usepackage{subfig}
\usepackage{float}
\usepackage{multicol}
\usepackage[usenames,dvipsnames]{pstricks}
\usepackage{epsfig}
\usepackage{pst-grad} 
\usepackage{pst-plot} 
\usepackage[noadjust]{cite}
\usepackage{caption}
\usepackage{graphicx}
\usepackage{epstopdf}
\usepackage{psfrag}
\usepackage{amsfonts}
\usepackage{color}
\usepackage{amsmath,mathtools}
\usepackage{psfrag}
\usepackage{comment}
\usepackage{url}
\usepackage{balance}

\begin{document}
\title{Impact of delayed acceleration feedback \\ on the classical car-following model}

\author{\IEEEauthorblockN{Gopal Krishna Kamath, Krishna Jagannathan and Gaurav Raina}
\IEEEauthorblockA{\\Department of Electrical Engineering, Indian Institute of Technology Madras, Chennai 600 036, India\\
Email: $\lbrace \text{ee12d033, krishnaj, gaurav} \rbrace$@ee.iitm.ac.in}
\thanks{This is an extension of our preliminary work that appeared in Proceedings of the 2016 IEEE Conference on Control Applications (CCA), pp. 1336-1343, 2016. DOI: 10.1109/CCA.2016.7587992}}
\maketitle

\begin{abstract}
Delayed feedback plays a vital role in determining the qualitative dynamical properties of a platoon of vehicles driving on a straight road. Motivated by the positive impact of Delayed Acceleration Feedback (DAF) in various scenarios, in this paper, we incorporate DAF into the Classical Car-Following Model (CCFM). We begin by deriving the Classical Car-Following Model with Delayed Acceleration Feedback (CCFM-DAF). We then derive the necessary and sufficient condition for local stability of the CCFM-DAF. Next, we show that the CCFM-DAF transits from the locally stable to the unstable regime via a Hopf bifurcation; thus leading to the emergence of limit cycles in system dynamics. We then propose a suitable linear transformation that enables us to analyze the local bifurcation properties of the CCFM-DAF by studying the analogous properties of the CCFM. We also study the impact of DAF on three important dynamical properties of the CCFM; namely, non-oscillatory convergence, string stability and robust stability. Our analyses are complemented with a stability chart and a bifurcation diagram. Our work reveals the following detrimental effects of DAF on the CCFM: $(i)$ reduction in the locally stable region, $(ii)$ increase in the frequency of the emergent limit cycles, $(iii)$ decrease in the amplitude of the emergent limit cycles, $(iv)$ destruction of the non-oscillatory property, $(vi)$ increased risk of string instability, and $(vii)$ reduced resilience towards parametric uncertainty. Thus, we report a practically-relevant application wherein DAF degrades the performance in several metrics of interest.
\end{abstract}


\IEEEpeerreviewmaketitle

\section{INTRODUCTION}
\label{section:Introduction}
Various forms of delays such as reaction delays, feedback delays and communication delays are known to affect several properties of a dynamical system in many different ways~\cite{HL}. For instance, delay-induced instabilities and oscillations have been widely studied in different scenarios~\cite{HL, RS, XZ, GKK, JI}. In contrast, delays are also known to aid stability of system, when introduced appropriately~\cite{JI, TI}.

In the context of car-following models -- a class of dynamical systems used to describe vehicular motion -- reaction delays usually have a detrimental effect. Specifically, increasing the reaction delay is known destabilize the resulting traffic flow~\cite{GKK, GO, arx}. Specifically, Kamath \emph{et al.}~\cite{GKK} showed that an increase in the reaction delay can lead to peculiar phenomenon known as a `phantom jam'~\cite{DC, DH} -- the emergence of back-propagating congestion waves in motorway traffic, seemingly out of nowhere. In contrast, Ge~\emph{et al.}~\cite{JI} showed that a specific car-following model (the Optimal Velocity Model~\cite{MBD}) became more resilient towards reaction and communication delays with the use of Delayed Acceleration Feedback (DAF) signal. Motivated by this, in this paper, we investigate the impact of DAF on the qualitative dynamical properties of a platoon of vehicles traversing a straight road with no overtaking. Specifically, we analyze the effects of DAF on the Classical Car-Following Model (CCFM)~\cite{arx}.

\subsection{Related work on car-following models and DAF}
Car-following models are a class of dynamical models that describe the temporal evolution of the position, velocity and acceleration of each vehicle in vehicular platoons. Chandler~\emph{et al.}~\cite{REC} and Herman~\emph{et al.}~\cite{RH} were some of the earliest to study the stability aspects of car-following models. Our work builds on the CCFM proposed in the pioneering work by Gazis~\emph{et al.}~\cite{DCG}. Several related models have been analyzed in~\cite{DCG},~\cite{EAU}, and~\cite{REC}. Wilson \emph{et al.}~\cite{REW} provide an exposition of linear stability analysis in the context of car-following models.

The above works predominantly make use of transform techniques to derive conditions for stability. In contrast, Zhang \emph{et al.}~\cite{XZ} and some of the references therein, treat the issue of stability from a dynamical systems perspective. From this viewpoint, Kamath~\emph{et al.}~\cite{GKK} proved that a version of the CCFM loses stability via a Hopf bifurcation. This, in turn, leads to the emergence of limit cycles (isolated closed orbits in phase space), which manifest as back-propagating congestion waves in motorway traffic.

Acceleration and DAF signals have been studied in many different applications~\cite{HG, TI, TV}. Specifically, in the context of human postural balance, Insperger~\emph{et al.}~\cite{TI} proposed a Proportional Derivative Acceleration (PDA) controller, which makes use of DAF. Therein, the authors numerically establish the superiority of the PDA over the classical proportional derivative controller. This motivated Ge~\emph{et al.}~\cite{JI} to make use of a variant of the PDA controller in the context of stabilizing a platoon of vehicles. To that end, the authors make use of appropriately-placed connected cruise control vehicles running the OVM. Therein, Ge~\emph{et al.} numerically showed that the PDA controller could increase the resilience of the system to reaction and communication delays.

Motivated by the work of Ge~\emph{et al.}, Kamath \emph{et al.} incorporated the DAF in a version of the CCFM (called the RCCFM -- Reduced Classical Car-Following Model)~\cite{multi}. In contrast to the literature, it was \emph{analytically} shown that DAF had a \emph{detrimental} effect of stability of the RCCFM. In this paper, we extend the results of~\cite{multi} to the CCFM, in addition to studying the effects of DAF on three additional important properties; namely, non-oscillatory convergence, string stability and robust stability.

\subsection{Our contributions}

Our contributions can be summarized as follows.
\begin{itemize}
\item[(1)] We derive the necessary and sufficient condition for local stability of the Classical Car-Following Model with Delayed Acceleration Feedback (CCFM-DAF) -- a model derived by incorporating DAF in the CCFM. From the resulting analysis, we show that DAF shrinks the locally stable region of the CCFM.
\item[(2)] We analytically prove that the CCFM-DAF also undergoes a Hopf bifurcation, thus resulting in the emergence of limit cycles in system dynamics. This shows that DAF preserves the manner in which the CCFM loses stability. We also show that DAF increases the frequency of the emergent limit cycles.
\item[(3)] We use a suitable linear transformation to deduce the type of Hopf bifurcation and the asymptotic orbital stability of the emergent limit cycles of the CCFM-DAF by analyzing the CCFM. That is, we provide a method to study local bifurcation properties of the CCFM-DAF by studying the analogous properties of the CCFM instead. This is of significance since the CCFM-DAF is governed by a system of neutral functional differential equations, whereas the CCFM is governed by a system of retarded functional differential equations, which are relatively easier to analyze.
\item[(4)] We show that DAF destroys the non-oscillatory property of the CCFM's solutions. This, in turn, means that achieving smooth traffic flow for the CCFM-DAF is impossible due to the presence of jerky vehicular motion.
\item[(5)] We derive a sufficient condition for string stability of the CCFM-DAF. We show that, with increase in the DAF signal strength, the corresponding region for string stability vanishes.
\item[(6)] We derive a necessary condition for robust stability of the CCFM-DAF to parametric uncertainties. We then show that the CCFM-DAF loses its resilience to parametric uncertainty as the DAF signal strength is increased.
\item[(7)] And lastly, we numerically show that DAF decreases the amplitude of limit cycles. Thus, on the boundary of stability, DAF may result in fast and short jerks in vehicular motion.
\end{itemize}

Thus, this paper reports a practically-relevant application wherein DAF is analytically shown to degrade the performance in several measures of interest.

The remainder of this paper is organized as follows. In Section~\ref{section:Models}, we present the scenario for our work, briefly describe the CCFM, and also derive the CCFM-DAF. In Section~\ref{section:HB}, we study the impact of DAF on local stability of the CCFM and also its loss of local stability. In Sections~\ref{section:NoC},~\ref{section:SS} and~\ref{section:RS}, we study the effect of DAF on non-oscillatory convergence, string stability and robust stability of the CCFM. In Section~\ref{section:NumCom}, we present a numerically-constructed bifurcation diagram, and conclude this paper in Section~\ref{section:Conclusions}.

\section{Models}
\label{section:Models}
In this section, we first describe the scenario for our work. We then briefly describe the CCFM. Finally, we derive the CCFM-DAF, which we study in the remainder of this paper.

\subsection{The setting}
In this paper, we assume a platoon of $N+1$ self-driven vehicles traveling on an infinitely long, single-lane road without overtaking. We assume that each vehicle is well modeled by a point, \emph{i.e.,} they have no length. The lead vehicle is indexed $0,$ the vehicle behind it by $1,$ and so forth. The acceleration of each vehicle is updated depending on a combination of its position, velocity and acceleration and also those of the vehicle directly ahead. We use $x_i(t)$ to denote the position of the $i^{th}$ vehicle at time $t.$ Hence, by standard convention, its velocity and acceleration at time $t$ will be denoted by $\dot{x}_i(t)$ and $\ddot{x}_i(t)$ respectively. We assume that the dynamics of the lead vehicle are known. That is, we assume that the variation in the acceleration and the velocity of the lead vehicle. In particular, we consider only those leader profiles that converge, in finite time, to $\ddot{x}_0 = 0$ and $\dot{x}_0 < \infty$; that is, there exists a finite $T_0$ such that $\ddot{x}_0(t) = 0,$ $\dot{x}_0(t) = \dot{x}_0,$ $\forall t \geq T_0$. We use the terms ``driver" and ``vehicle" interchangeably. We also use SI units throughout.

\subsection{The Classical Car-Following Model (CCFM)}
The update equations for the CCFM are given by~\cite{DCG}
\begin{align}
\label{eq:CCFMO}
\ddot{x}_i(t) = \alpha_i \frac{ \left( \dot{x}_i(t) \right)^m \left( \dot{x}_{i-1}(t - \tau) - \dot{x}_{i}(t - \tau) \right)}{\left( x_{i-1}(t - \tau) - x_{i}(t - \tau)\right)^l},
\end{align}
for $i \in \lbrace 1, 2, \cdots, N \rbrace$. Here, $\alpha_i > 0$ represents the $i^{th}$ driver's sensitivity coefficient. Also, $m \in [-2,2]$ and $l \in \mathbb{R}_{+}$ are model parameters that contribute to the non-linearity.

Following~\cite{arx}, we set $y_i(t)$ + $b_i$ = $x_{i-1}(t) - x_i(t)$ and $v_i(t)$ = $\dot{y}_i(t)$ = $\dot{x}_{i-1}(t) - \dot{x}_i(t).$ We also account for the delay in the self-velocity term, as in~\cite{arx}. Thus, system~\eqref{eq:CCFMO} becomes:
\begin{align}
\nonumber
\dot{v}_i(t) = & \, \beta_{i-1}(t - \tau_{i-1}) v_{i-1}(t - \tau_{i-1}) - \beta_i( t - \tau_i)  v_{i}(t - \tau_i), \\ \label{eq:CCFMT12}
\dot{y}_i(t) = & \, v_i(t),
\end{align}
for $i \in \lbrace 1, 2, \cdots, N \rbrace.$ Here,
\begin{align*}
\beta_i(t) = \, \alpha_{i} \frac{ \left( \dot{x}_0(t) - v_0(t) - \cdots - v_{i}(t) \right)^m}{\left( y_{i}(t) + b_i \right)^l}.
\end{align*}
Here, $b_i$ denotes the desired equilibrium separation of the $i^{th}$ vehicle, and $y_i(t) + b_i$ represents the separation between vehicles $i-1$ and $i$ at time $t.$ Thus, $y_i(t)$ denotes the variation of the headway about its desired values $b_i,$ at time $t.$ Further, $v_i(t)$ corresponds to the relative velocity of the $i^{th}$ vehicle with respect to the $(i-1)^{th}$ vehicle at time $t.$ Note that $y_0$, $v_0$, $\alpha_0$ and $\tau_0$ are dummy variables introduced for notational brevity, all of which are set to zero. We emphasize that $y_0$ and $v_0$ are \emph{not} state variables.

\subsection{The Classical Car-Following Model with Delayed Acceleration Feedback (CCFM-DAF)}

We now derive the CCFM-DAF, obtained by incorporating DAF in the evolution equations of the CCFM. We begin with~\eqref{eq:CCFMO} and introduce a delayed acceleration term, similar to~\cite{TI}, to obtain

\vspace*{-3mm}
\begin{small}
\begin{align}
\label{eq:CCFMOl0wda}
\ddot{x}_i(t) = \, \alpha_i \frac{ \left( \dot{x}_i(t) \right)^m \left( \dot{x}_{i-1}(t - \tau) - \dot{x}_{i}(t - \tau) \right)}{\left( x_{i-1}(t - \tau) - x_{i}(t - \tau)\right)^l} + \gamma \ddot{x}_i(t - \tau),
\end{align}
\end{small}

\vspace*{-4mm}
\noindent for $i \in \lbrace 1, 2, \cdots, N \rbrace.$ Here, $\gamma > 0$ captures the sensitivity towards delayed acceleration. Transforming~\eqref{eq:CCFMOl0wda} similar to the CCFM, accounting for the heterogeneity in reaction delays and sensitivity coefficients, and re-arranging the terms, we obtain the following system:
\begin{align}
\nonumber
\dot{v}_1(t) - \gamma_1 \dot{v}_1(t - \tau_1) = & \, \ddot{x}_0(t) - \gamma_1 \ddot{x}_0(t - \tau_1) - \beta_{1}(t - \tau_1) v_{1}(t - \tau_1),  \\  \label{eq:RCCFMTwDAF}
\dot{v}_k(t) - \gamma_k \dot{v}_k(t - \tau_k)  = & \, \beta_{k-1}(t - \tau_{k-1}) v_{k-1}(t - \tau_{k-1}) - \beta_{k}(t - \tau_k) v_{k}(t - \tau_k),
\end{align}
for $k$ $\in$ $\lbrace 2, 3, \cdots, N \rbrace$ and $\beta_i(t)$ as in the CCFM. We refer to system~\eqref{eq:RCCFMTwDAF} as the Classical Car-Following Model with Delayed Acceleration Feedback (CCFM-DAF). Note that system~\eqref{eq:RCCFMTwDAF} is a system of neutral functional differential equations~\cite[Section 2.7]{HL}, since the highest order derivative of the state variable is delayed. Also note that the delay incurred while sensing the acceleration signal is negligible. However, to keep the analysis general, we do not make use of any approximations.

\section{The Hopf Bifurcation}
\label{section:HB}
Hopf bifurcation~\cite{HL} is a phenomenon wherein a dynamical system loses stability due to a pair of conjugate eigenvalues crossing the imaginary axis in the Argand plane. Mathematically, a Hopf bifurcation analysis is a rigorous way of proving the emergence of limit cycles in non-linear dynamical systems.

In this section, study the bifurcation properties of the CCFM-DAF. We begin by linearizing the CCFM-DAF described by~\eqref{eq:RCCFMTwDAF}. We then derive the necessary and sufficient condition for local stability of the CCFM-DAF. We then show that the transversality condition of the Hopf spectrum~\cite[Chapter 11, Theorem 1.1]{HL} holds for the CCFM-DAF. Thus, we conclude that the CCFM-DAF loses local stability via a Hopf bifurcation, thereby leading to the emergence of limit cycles in system dynamics. We then use a suitable linear transformation to obtain insight into local bifurcation properties of the CCFM-DAF by analyzing the CCFM. This provides a way of analyzing a class of neutral functional differential equations by analyzing the associated retarded functional differential equation.

\subsection{Transversality condition of the Hopf spectrum}
\label{section:RCCFMwDFALSA}


We begin by linearizing system~\eqref{eq:RCCFMTwDAF} about a desired equilibrium. To that end, note that $v_i^* = 0,$ $y_i^* = 0,$ $i = 1, 2, \cdots, N$ is an equilibrium for system~\eqref{eq:RCCFMTwDAF}. Linearizing system~\eqref{eq:RCCFMTwDAF} about this equilibrium, and setting the leader's profile to zero, we obtain
\begin{align*}
\dot{v}_i(t) = & \, \beta_{i-1}^* v_{i-1}(t - \tau_{i-1}) - \beta_i^* v_{i}(t - \tau_{i}) + \gamma_i \dot{v}_i(t - \tau_i), \\
\dot{y}_i(t) = & \, v_i(t),
\end{align*}
for $i \in \lbrace 1, 2, \cdots, N \rbrace.$ Here, $\beta_i^* = \alpha_{i} (\dot{x}_0)^m / (b_i)^l$ denotes the equilibrium coefficient for the $i^{th}$ vehicle. Note that, in the vicinity of the equilibrium, the evolution of $v_i$ does not depend on $y_i.$ Hence, similar to~\cite{arx}, we drop the variables $\lbrace y_i \rbrace_{i = 1}^N.$ Thus, we obtain
\begin{align}
\label{eq:LRCCFMTwDFA}
\dot{v}_i(t) = & \, \beta_{i-1}^* v_{i-1}(t - \tau_{i-1}) - \beta_i^* v_{i}(t - \tau_{i}) + \gamma_i \dot{v}_i(t - \tau_i),
\end{align}
for $i \in \lbrace 1, 2, \cdots, N \rbrace.$ These equations can be succinctly written using matrix representation as
\begin{align}
\label{eq:CELRCFFMwDFA}
\dot{s}(t) = \sum\limits_{k = 1}^{N} \big( A_k s(t - \tau_k) + B_k \dot{s}(t - \tau_k)\big),
\end{align}
where $s(t) = [v_1(t) \text{ } v_2(t) \text{ } \cdots \text{ } v_N(t)]^T$ and $A_k, B_k \in \mathbb{R}^{N \times N}$ $\forall k.$ In fact, for $k \in \lbrace 1, 2, \cdots, N \rbrace,$
\begin{align*}
(B_k)_{ij} = \begin{cases}
\gamma_k , & i = j = k,\\
0, & \text{otherwise}.
\end{cases}
\end{align*}
Also, for $k \in \lbrace 1, 2, \cdots, N-1 \rbrace,$
\begin{align*}
(A_k)_{ij} = \begin{cases}
-\beta_k^{*} , & i = j = k,\\
\beta_k^{*} , & j = k, i = k+1,\\
0, & \text{otherwise},
\end{cases}
\end{align*}
and
\begin{align*}
(A_N)_{ij} = \begin{cases}
-\beta_N^{*} , & i = j = k,\\
0, & \text{otherwise}.
\end{cases}
\end{align*}

The characteristic equation of~\eqref{eq:CELRCFFMwDFA}, is given by~\cite[Section 5.1]{GL}
 $$f(\lambda) = \text{det} \left(\lambda I_{N \times N} - \sum\limits_{k = 1}^{N} (e^{- \lambda \tau_k} A_k + \lambda e^{- \lambda \tau_k} B_k) \right) = 0.$$
On simplification, this yields
\begin{align}
\label{eq:CELRCFFMwDAF}
f(\lambda) = \, \overset{N}{\underset{i = 1}{\prod}} (\lambda - \gamma_i \lambda e^{- \lambda \tau_i} + \beta_i^* e^{- \lambda \tau_i}) = 0.
\end{align}
Typically, the above term is zero due to exactly one term being zero. Therefore, for some $i \in \lbrace 1, 2, \cdots, N \rbrace$, we have
\begin{align}
\label{eq:RCELRCFFMwDAF}
\lambda - \gamma_i \lambda e^{- \lambda \tau_i}  + \beta_i^* e^{- \lambda \tau_i} = \, 0.
\end{align}

Note that~\eqref{eq:RCCFMTwDAF} is a system of neutral-type equations. Therefore, we require an addition `neutral condition' for the system stability to be characterized by its eigenvalues being located in the open left half of the Argand plane. For systems of the form~\eqref{eq:RCCFMTwDAF}, this condition is given by $\gamma_i < 1$~\cite[Section 1.7]{HL}. Therefore, we enforce this constraint in the remainder of this paper. We now search for a conjugate pair of eigenvalues of~\eqref{eq:RCELRCFFMwDAF} crossing the imaginary axis in the Argand plane, thereby pushing the system into an unstable regime. To that end, we substitute $\lambda = j \omega$, where $j = \sqrt{-1}$, in~\eqref{eq:RCELRCFFMwDAF} to obtain
\begin{align*}
& \beta_i^* \text{ cos}(\omega \tau_i) - \gamma_i \omega \text{ sin}(\omega \tau_i) = 0, \\
& \omega - \gamma_i \omega \text{ cos}(\omega \tau_i) - \beta_i^* \text{ sin}(\omega \tau_i) = 0.
\end{align*}
From the first equality, we have $\beta_i^* = \gamma_i \omega$ tan$(\omega \tau_i).$ Using this, the second equality can then be simplified to $\text{ cos}(\omega \tau_i) = \gamma_i$ and $\text{ sin}(\omega \tau_i) = \beta_i^*/\omega.$ From these, the angular velocity of the oscillations is obtained as
\begin{align*}
\omega_0 = \frac{\beta_i^*}{\sqrt{1 - \gamma_i^2}}.
\end{align*}
Note that, due to the neutral condition ($\gamma_i < 1$), we have $\omega_0 > 0.$ Therefore, when exactly one conjugate pair of eigenvalues lies on the imaginary axis in the Argand plane, we have
\begin{align}
\label{eq:RCCFMONwDAF}
\omega_0 = & \, \frac{\beta_i^*}{\sqrt{1 - \gamma_i^2}}, \\ \label{eq:RCCFMKCwDAF}
\tau_{i_{cr}} = & \, \frac{1}{\omega_0} \text{tan}^{-1}\left( \frac{\beta_i^*}{\gamma_i \omega_0} \right),
\end{align}
where $\tau_{i_{cr}}$ denotes the critical value of the reaction delay when $\omega = \omega_0.$

We now prove the transversality condition of the Hopf spectrum~\cite[Chapter 11, Theorem 1.1]{HL}
\begin{align}
\label{eq:HopfTransCond}
\text{Real}\left( \frac{\text{d} \lambda}{\text{d} \tau_i} \right)_{\tau_i = \tau_{i_{cr}} } \neq \, 0.
\end{align}.
This would then imply that the CCFM-DAF undergoes a Hopf bifurcation at $\tau_i = \tau_{i_{cr}}.$ To that end, we differentiate~\eqref{eq:RCELRCFFMwDAF} with respect to $\tau_i,$ and simplifying, to obtain
\begin{align}
\label{eq:RCCFMtranscondwDAF}
\text{Real}\left( \frac{\text{d} \lambda}{\text{d} \tau_i} \right)_{\tau_i = \tau_{i_{cr}} } = \frac{\omega_0^2 (1 - \gamma_i^2 )}{\tilde{\theta}} > 0,
\end{align}
where $\tilde{\theta} = (1 - \gamma_i \text{ cos}(\omega_o \tau_i))^2 + (\omega_0 \tau_i + \gamma_i \text{ sin}(\omega_0 \tau_i))^2.$ Note that the neutral condition $\gamma_i < 1$ ensures the positivity of the Right Hand Side (RHS) above. Therefore, system~\eqref{eq:RCCFMTwDAF} undergoes a Hopf bifurcation at $\tau_i = \tau_{i_{cr}}.$ Hence, $\tau_i < \tau_{i_{cr}}$ is the necessary and sufficient condition for system~\eqref{eq:RCCFMTwDAF} to be locally stable when the neutral condition $\gamma_i < 1$ is satisfied. Further, when $\gamma_i < 1$ holds, $\tau_i = \tau_{i_{cr}}$ represents the equation of the stability boundary, also known as the Hopf boundary.

\begin{figure}[tbh]
\centering
\psfrag{0.0}{\scriptsize \hspace*{1mm} $0$}
\psfrag{0.2}{\scriptsize \hspace*{1mm} $0.2$}
\psfrag{0.4}{\scriptsize \hspace*{1mm} $0.4$}
\psfrag{0.6}{\scriptsize \hspace*{1mm} $0.6$}
\psfrag{0.8}{\scriptsize \hspace*{1mm} $0.8$}
\psfrag{1.0}{\scriptsize \hspace*{1mm} $1$}
\psfrag{1.2}{\scriptsize $1.2$}
\psfrag{1.6}{\scriptsize $1.6$}
\psfrag{t}{\small $\tau_{i_{cr}}$}
\psfrag{g}{\small $\gamma_i$}
\includegraphics[scale=0.45,angle=-90]{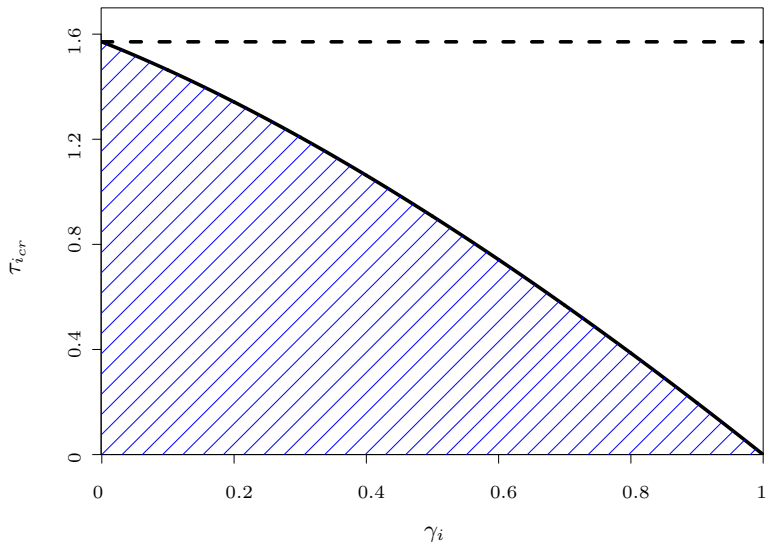}
\caption{\emph{Stability chart:} The shaded region corresponds to the locally stable region for CCFM-DAF system satisfying $\beta_i^* = 1.$ Black solid line captures the decreasing trend of the left hand side of~\eqref{eq:comparison} with increase in $\gamma_i.$ The dashed line represents $\pi/2,$ for reference.}
\end{figure}

We now infer the effect of DAF on some local properties of the CCFM. We begin by comparing the critical values of the reaction delay under the CCFM-DAF and the CCFM. To that end, we simplify~\eqref{eq:RCCFMKCwDAF}, to obtain
\begin{align}
\label{eq:taucritDAF}
\tau_{i_{cr}} = \frac{\sqrt{1 - \gamma_i^2}}{\beta_i^*} \text{ tan}^{-1}\left(\frac{\sqrt{1 - \gamma_i^2}}{\gamma_i}\right).
\end{align}
In contrast, the critical value of the reaction delay under the CCFM, to be denoted by $\tilde{\tau}_{i_{cr}},$ was shown to be~\cite[Equation~$(21)$]{arx}
\begin{align}
\label{eq:taucritRCCFM}
\tilde{\tau}_{i_{cr}} = \frac{\pi}{2\beta_i^*}.
\end{align}
For the DAF to have a positive impact on the locally stable region of the CCFM, it suffices to show that
\begin{align}
\label{eq:comparison}
\sqrt{1 - \gamma_i^2} \text{ tan}^{-1}\left(\frac{\sqrt{1 - \gamma_i^2}}{\gamma_i}\right) \geq \frac{\pi}{2},
\end{align}
is true. This would ensure that DAF has enlarged (in the set containment sense) locally stable region in the parameter space. Note that~\eqref{eq:comparison} is trivially met with equality for $\gamma_i = 0,$ whence system~\eqref{eq:RCCFMTwDAF} collapses to the CCFM. Fig.~$1$ portrays the variation of the left hand side of~\eqref{eq:comparison} as $\gamma_i$ is increased. From this, we can infer the non-existence of a non-trivial $\gamma_i$ satisfying~\eqref{eq:comparison}. That is, DAF is detrimental to the stability of the CCFM. In fact, Fig.~$1$ represents the stability chart for CCFM-DAF systems with $\beta_i^* = 1.$ Note that an increase in $\gamma_i$ results in destabilizing of the system for a relatively smaller value of the reaction delay. thereby decreasing the resilience of the CCFM.

\subsection{Local bifurcation properties of the CCFM-DAF}

Thus far, we showed that the CCFM-DAF loses local stability via a Hopf bifurcation, thus leading to the emergence of limit cycles in system dynamics. The next natural step is to characterize the \emph{type} of the Hopf bifurcation and the asymptotic \emph{orbital stability} of the resulting limit cycles. Such an analysis for neutral-type equations is, in general, relatively harder than for their retarded-type counterparts. Thus, in this sub-section, we propose a method of analyzing local bifurcation properties of the CCFM-DAF by analyzing their counterparts for the CCFM.

For notational convenience, we denote $v_i(t) - \gamma_i v_i(t - \tau_i)$ by $l_i(t)$ in~\eqref{eq:RCCFMTwDAF}. Hence, the RHS of~\eqref{eq:RCCFMTwDAF} captures the dynamics of $l_i(t).$ In other words, for $i \in \lbrace 1, 2, \cdots, N \rbrace,$ we have

\vspace*{-3mm}
\begin{small}
\begin{align}
\label{eq:modelrhs}
\dot{l}_i(t) = \beta_{i-1}(t - \tau_{i-1}) \text{ } v_{i-1}(t - \tau_{i-1}) - \beta_{i}(t - \tau_i) \text{ } v_{i}(t - \tau_{i}).
\end{align}
\end{small}
\vspace*{-3mm}

\noindent Note that the variable $l_i(t)$ can be expressed as $l_i(t) = \langle c,\textbf{v}(t) \rangle = [1 \text{ } -\gamma_i] [v_i(t) \text{ } v_i(t - \tau_i)]^T,$ where $c = [1 \text{ } -\gamma_i]^T$ and $\textbf{v}(t) = [v_i(t) \text{ } v_i(t - \tau_i)]^T.$ We now prove a general result, which we then use for our specific system.

\emph{Lemma~$1$:} Let $x(t) = [x_1(t) \text{ } x_2(t)]^T$ and $c = [c_1 \text{ } c_2]^T,$ where $x_1$ and $x_2$ are bounded, non-constant real-valued functions and $c_1$ and $c_2$ are non-zero, real constants. Also, let $y(t) = \langle c,x(t) \rangle = c_1 x_1(t) + c_2 x_2(t).$ Then, $y(t)$ is periodic if and only if $x(t)$ is periodic. Moreover, $x(t)$ and $y(t)$ will have the same period.

\emph{Proof:} First, let $x(t)$ be periodic with period $T > 0,$ $i.e.,$ $x(t+T) = x(t)$ $\forall t.$ Then, $y(t+T) = \langle c,x(t+T) \rangle = \langle c,x(t) \rangle = y(t)$ $\forall t.$ Hence, $y(t)$ is periodic with period $T.$ Conversely, assume that $y(t)$ is periodic with period $T > 0,$ $i.e.,$ $y(t+T) = y(t)$ $\forall t.$ Then, $\langle c,x(t+T)  \rangle = \langle c,x(t)  \rangle$ $\forall$ $t.$ Therefore, $\langle c, x(t) - x(t+T) \rangle = 0$ $\forall$ $t.$ Since $c_1$ and $c_2$ are non-zero, the assumptions on $x_1$ and $x_2$ imply that $x(t) = x(t+T)$ $\forall$ $t,$ $i.e.,$ $x(t)$ is periodic with period $T.$ \hspace*{4mm} $\square$

In the context of the CCFM-DAF, Lemma~$1$ implies that $v_i(t)$ is periodic if and only if  $l_i(t)$ is periodic, and that their periods coincide. Moreover, this equivalence also implies that the aforementioned transformation preserves the topological changes in the phase space, $i.e.,$ $v_i(t)$ would undergo the same type of Hopf bifurcation as $l_i(t)$ and the emergent limit cycles would have the same asymptotic orbital stability. Note that the detailed Hopf bifurcation analysis for the CCFM has been performed in~\cite[Section~VIII]{arx}.  In the next sub-section, we discuss the effect of DAF on the frequency of emergent limit cycles. In Section~\ref{section:NumCom}, we will numerically evaluate how DAF affects the amplitude of emergent limit cycles.

\subsection{Discussion}
\label{section:disc}

A few remarks are in order. 
\begin{itemize}
\item[(1)] Note that the system loses stability when the very first conjugate pair of eigenvalues crosses the imaginary axis in the Argand plane. Since the derivative~\eqref{eq:RCCFMtranscondwDAF} is positive, increasing the reaction delay beyond its critical value cannot restore stability. Specifically, the CCFM-DAF cannot undergo \emph{stability switches}.
\item[(2)] From~\eqref{eq:RCCFMKCwDAF}, note that the critical value of the reaction delay $\tau_{i_{cr}}$ depends on system parameters. This implies that an appropriate variation in \emph{any} of these parameters could push the system into the unstable regime. Thus, we can choose any of these parameters to act as the bifurcation parameter. Alternatively, we could even use an exogenous non-dimensional parameter as the bifurcation parameter, as done in~\cite{GKK, arx}.
\item[(3)] By $f_0$ and $\tilde{f}_0,$ we denote the frequency of the emergent limit cycles of the CCFM-DAF and the CCFM respectively. Then, from~\eqref{eq:RCCFMONwDAF}, we have $f_0 = \omega_0 / 2\pi.$ Similarly, from~\cite[Equation~$(21)$]{arx}, we have $\tilde{f}_0 = \beta_i^* / 2\pi.$ Comparing the two, we have $f_0 \geq \tilde{f}_0.$ Therefore, we infer that DAF increases the frequency of the emergent limit cycles of the CCFM. In summary, in addition to shrinking the locally stable region of the CCFM, DAF increases the frequency of emergent limit cycles. In the context of transportation networks, this is detrimental since oscillations of increased frequency manifest as fast jerky vehicular motion; this may degrade the ride quality.
\item[(4)] Finally, note that Olfati-Saber \emph{et al.} derive an expression similar to~\eqref{eq:taucritRCCFM} in the context of consensus dynamics~\cite{ROS}. Hence, our results apply to \emph{any} system with a characteristic equation similar to~\eqref{eq:RCELRCFFMwDAF}, independent of the underlying application.
\end{itemize}

\section{Non-oscillatory convergence}
\label{section:NoC}
In this section, we study the effect of DAF on the non-oscillatory property of solutions of the CCFM. Since relative velocities and headways form the dynamical variables for the CCFM-DAF, it is desirable for the solutions to possess the non-oscillatory property. This may ensure the resulting traffic flow to be smooth and the ride quality good, due to lack of jerky vehicular motion.

Note that the solutions of the CCFM possess the non-oscillatory property if and only if the parameters satisfy~\cite[Equation $(23)$]{arx}
\begin{align*}
\beta_i^* \tau_i \leq \frac{1}{e},
\end{align*}
for $i \in \lbrace 1, 2, \cdots, N \rbrace.$ To understand the effect of DAF, we re-state a version of~\cite[Theorem 6.1.3]{GL} as follows.

\emph{Lemma~$2$:} Consider a system with the characteristic equation
\begin{align}
\label{eq:glce}
f(\lambda) = \lambda + p \lambda e^{-\lambda \tau} + q e^{-\lambda \sigma} = 0,
\end{align}
where $p \in \mathbb{R},$ $\tau, q \in (0, \infty)$ and $\sigma \in [0, \infty).$ Then, every solution of the said system oscillates if
\begin{align}
\label{eq:glsc}
p \neq -1 \text{ and } \frac{q(\sigma - \tau)}{1 + p} < \frac{1}{e}.
\end{align}

Comparing the characteristic equation of the CCFM-DAF~\eqref{eq:RCELRCFFMwDAF} with~\eqref{eq:glce}, we obtain $p = - \gamma_i,$ $\tau = \sigma = \tau_i$ and $q = \beta_i^*.$ Due to the neutral condition $\gamma_i < 1,$ we have $p \neq -1$ being automatically satisfied. Similarly, since $\tau = \sigma,$ the inequality in~\eqref{eq:glsc} is also satisfied. Thus, the CCFM-DAF satisfies the sufficient condition of Lemma~$2.$ Therefore, \emph{every} solution of the CCFM-DAF oscillates. In other words, DAF destroys the non-oscillatory property of the solutions of the CCFM. Therefore, incorporating DAF may result in persistent jerky vehicular motion, thereby degrading ride quality.

\section{String stability}
\label{section:SS}
It is well known that ensuring pairwise stability of vehicles does not guarantee that the entire platoon will be stable~\cite{LP}. Hence, we need to also ensure \emph{string stability, i.e.,} we need to ensure that the disturbance in state variables does not propagate along the platoon. In this section, we derive a sufficient condition for string stability of the CCFM-DAF, following closely the methodology used in~\cite{ifac}; we ensure that the magnitude-squared Bode plot of the pairwise interaction lies below unity for all frequencies.

We begin by noting that the dynamics of pairwise evolution of state variables for the CCFM-DAF is given by
\begin{align*}
\dot{v}_i(t) - \gamma_i \dot{v}_i(t - \tau_i)  = & \, \beta_{i-1}(t - \tau_{i-1}) v_{i-1}(t - \tau_{i-1}) - \beta_{i}(t - \tau_i) v_{i}(t - \tau_i),
\end{align*}
for $i \in \lbrace 1, 2, \cdots, N \rbrace.$ Taking Laplace transform and simplifying, we obtain the transfer function of pairwise interaction as
\begin{align*}
H_i(s) = \frac{V_{i}(s)}{V_{i-1}(s)} = \frac{\beta_{i-1}^* e^{- s \tau_{i-1}}}{s + (\gamma_i s + \beta_i^*) e^{-s \tau_i}}.
\end{align*}
To obtain the magnitude-squared Bode plot, we substitute $s = j \omega$ and simplify. This yields
\begin{align}
\label{eq:ssbp}
|H_i(j \omega)|^2 = \frac{(\beta_{i-1}^*)^2}{\omega^2 (1 + \gamma_i^2 + 2 \gamma_i \cos(\omega \tau_i)) - 2 \beta_i^* \omega \sin(\omega \tau_i)  + (\beta_{i}^*)^2}.
\end{align}

Note that $\omega^2 + (\beta_{i}^*)^2 - 2 \beta_i^* \omega \sin(\omega \tau_i)$ $\geq$ $\omega^2 (1 + \gamma_i^2 + 2 \gamma_i \cos(\omega \tau_i)) - 2 \beta_i^* \omega \sin(\omega \tau_i)  + (\beta_{i}^*)^2.$ Therefore, we have
\begin{align*}
|H_{l_i}(j \omega)|^2 & = \frac{(\beta_{i-1}^*)^2}{(\omega + \beta_i^*)^2} \\
& \leq \frac{(\beta_{i-1}^*)^2}{\omega^2 - 2 \beta_i^* \omega \sin(\omega \tau_i)  + (\beta_{i}^*)^2} \leq |H_i(j \omega)|^2.
\end{align*}
Note that $|H_{l_i}(j \omega)|^2$ is a monotonic decreasing function in $\omega$ over $\omega \geq 0.$ Therefore, it suffices to ensure $|H_{l_i}(0)|^2 \leq 1.$ This mandates that $\beta_{i-1}^* \leq \beta_{i}^*$ hold true for each $i.$ Since $|H_{l_i}(j \omega)|^2$ is a lower bound on $|H_{i}(j \omega)|^2,$ $\beta_{i-1}^* \leq \beta_{i}^*$ for each $i$ is a necessary condition for string stability of the CCFM-DAF. In other words, if $\beta_{i-1}^* > \beta_{i}^*$ for some $i,$ then CCFM-DAF cannot be string stable. Therefore, we henceforth impose $\beta_{i-1}^* \leq \beta_{i}^*$ for each $i.$ Thus, we have
\begin{align}
\label{eq:ubss}
|H_i(j \omega)|^2 \leq \frac{(\beta_{i}^*)^2}{\omega^2 (1 + \gamma_i^2 + 2 \gamma_i \cos(\omega \tau_i)) - 2 \beta_i^* \omega \sin(\omega \tau_i)  + (\beta_{i}^*)^2}.
\end{align}
We now derive a condition for the upper bound to $|H_i(j \omega)|^2$ in~\eqref{eq:ubss} to be less than unity. This would yield a sufficient condition for string stability of the CCFM-DAF. To that end, note that the upper bound is less than or equal to $1$ if and only if $\omega^2 (1 + \gamma_i^2 + 2 \gamma_i \cos(\omega \tau_i)) - 2 \beta_i^* \omega \sin(\omega \tau_i)$ $\geq 0$ $\forall$ $\omega \geq 0.$ Re-arranging, we need to find a condition such that
\begin{align*}
\frac{2 \gamma_i \cos(\omega \tau_i) + 1 + \gamma_i^2}{2 \beta_i^* \tau_i} \geq \frac{\sin(\omega \tau_i)}{\omega \tau_i}
\end{align*}
holds true. Note that the RHS of the above equation is upper bounded by unity. Hence, to obtain a sufficient condition, we replace the RHS with $1.$ Further, note that $\cos(\omega \tau_i) \geq -1$ $\forall$ $\omega, \tau_i.$ Substituting these, we obtain
\begin{align*}
\beta_i^* \tau_i \leq \frac{(1 - \gamma_i)^2}{2}.
\end{align*}
That is, a sufficient condition for string stability of the CCFM-DAF is
\begin{align}
\label{eq:sssuff}
\beta_{i-1}^* \leq \beta_i^* \text{ and } \beta_i^* \tau_i \leq \frac{(1 - \gamma_i)^2}{2},
\end{align}
for $i \in \lbrace 1, 2, \cdots, N \rbrace.$

Note that~\eqref{eq:sssuff} reduces to~\cite[Equation $(8)$]{ifac} when $\gamma_i = 0.$ That is, we recover the sufficient condition for the CCFM from~\eqref{eq:sssuff} when $\gamma_i = 0.$ Further, as $\gamma_i$ increases to $1,$ the RHS of~\eqref{eq:sssuff} decreases. This shows a degradation in the region corresponding to string stability as the DAF signal strength is increased.

\section{Robust stability}
\label{section:RS}
In this section, we study the effect of DAF on the resilience of the CCFM to parametric uncertainty. To that end, we assume that the parameter $\beta_i^*$ is uncertain in an interval $[\underline{\beta}_i^*,\overline{\beta}_i^*],$ for each $i.$ Specifically, we assume that these parameters are time-invariant, \emph{i.e.,} their value is apriori unknown. However, once these values materialize in the said interval, they are fixed. We then derive a necessary condition for robust stability of the CCFM-DAF, and substitute the ``worst-case'' value of the uncertain parameter in this expression to infer the effect of DAF on the resilience of the CCFM to parametric uncertainty.

We begin by noting that the necessary and sufficient condition for local stability of the CCFM-DAF, which was derived in Section~\ref{section:HB}, is
\begin{align*}
\tau_i < \frac{\sqrt{1 - \gamma_i^2}}{\beta_i^*} \text{ tan}^{-1}\left(\frac{\sqrt{1 - \gamma_i^2}}{\gamma_i}\right),
\end{align*}
for $i \in \lbrace 1, 2, \cdots, N \rbrace.$ However, $\tan^{-1}(x) \leq \pi / 2$ $\forall$ $x \in \mathbb{R}.$ Therefore, a \emph{necessary} condition for local stability is
\begin{align*}
\tau_i < \frac{\pi \sqrt{1 - \gamma_i^2}}{2 \beta_i^*},
\end{align*}
for $i \in \lbrace 1, 2, \cdots, N \rbrace.$ Substituting the ``worst-case'' value of $\beta_i^*$ in this condition, we obtain a necessary condition for robust stability of the CCFM-DAF as
\begin{align}
\label{eq:rsnec}
\tau_i < \frac{\pi \sqrt{1 - \gamma_i^2}}{2 \overline{\beta}_i^*},
\end{align}
for $i \in \lbrace 1, 2, \cdots, N \rbrace.$

Note that as the DAF signal strength increases (\emph{i.e.,} as $\gamma_i$ increases to $1$), the RHS of~\eqref{eq:rsnec} decreases to zero. That is, the CCFM becomes less resilient to parametric uncertainty as the DAF signal strength increases.

\section{Numerical Computations}
\label{section:NumCom}
\begin{figure}[htb]
\centering
\psfrag{0.0}{\scriptsize \hspace*{1mm} $0$}
\psfrag{-0.2}{\scriptsize $-0.2$}
\psfrag{0.2}{\scriptsize $0.2$}
\psfrag{-0.4}{\scriptsize $-0.4$}
\psfrag{0.4}{\scriptsize $0.4$}
\psfrag{0.6}{\scriptsize $0.6$}
\psfrag{0.9}{\scriptsize $0.9$}
\psfrag{1.0}{\scriptsize \hspace*{0mm} $1$}
\psfrag{1.1}{\scriptsize $1.1$}
\psfrag{nnnnnnnnnn}{\small $\gamma_2 = 0.5$}
\psfrag{ssssssssss}{\small $\gamma_2 = 0.9$}
\psfrag{mmmmmmmmmm}{\small $\gamma_2 = 0$ (CCFM)}
\psfrag{t}{\small $\tau_{2}$}
\psfrag{amplitude}{\small \hspace*{7mm} Envelope of the emergent limit cycles}
\includegraphics[scale=0.45,angle=-90]{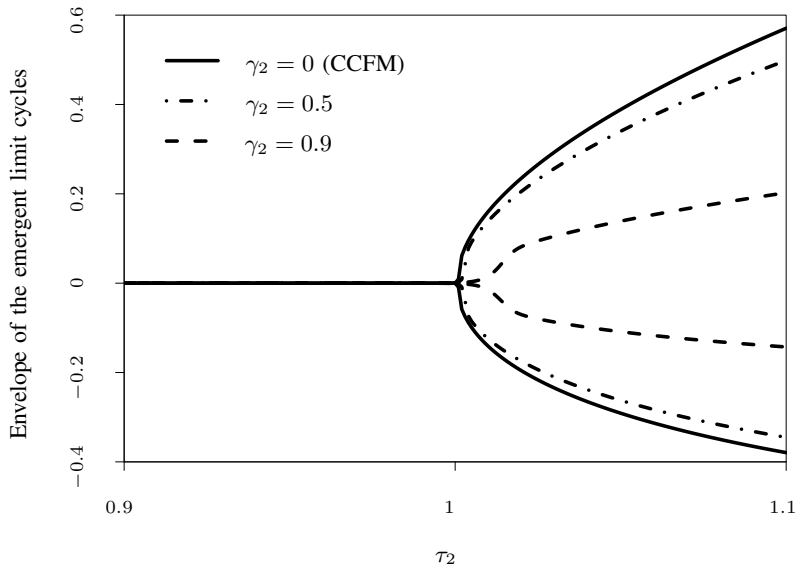}
\caption{\emph{Bifurcation diagram:} Plot of variation in the envelope of the emergent limit cycles with respect to $\tau_2,$ as $\gamma_2$  is varied in $\lbrace 0, 0.5, 0.9 \rbrace.$}
\end{figure}

In this section, we present a numerically-constructed bifurcation diagram to understand the effect of DAF on the amplitude of emergent limit cycles. We make use of the scientific computation software MATLAB to solve system~\eqref{eq:RCCFMTwDAF} using the `method of steps'~\cite[Chapter 5]{RDD}. We then plot the envelope (\emph{i.e.,} the maximum and the minimum values) of the resulting steady-state solution.

For the numerical computations, we consider $4$ vehicles following a lead vehicle, \emph{i.e.,} $N = 4,$ and assume that the $3^{rd}$ vehicle undergoes a Hopf bifurcation. We initialize the remaining parameter values to: $\alpha_1 = 0.1,$ $\alpha_2 = 0.5,$ $\alpha_3 = 0.3,$ $\tau_1 = 0.6,$ $\tau_2 = 1,$ $\tau_3= 0.2,$ $m = 1.5.$ We also assume uniform traffic flow with all headways of unit length. We then fix $\dot{x}_0$ using~\eqref{eq:RCCFMKCwDAF} to ensure $\tau_{2_{cr}} = 1,$ for simplicity. We then vary $\gamma_2$ in the range $\lbrace 0, 0.5, 0.9 \rbrace$ to infer the effect of DAF on the amplitude of the emergent limit cycles. From Fig.~$2,$ note the emergence of limit cycles when $\tau_2 = 1,$ as predicted by the analysis, as the system undergoes a Hopf bifurcation. Also note that increasing the DAF signal strength reduces the amplitude of the emergent limit cycles. Combining this observation with the discussion in Section~\ref{section:disc}, DAF leads to limit cycles of higher frequency and smaller amplitude. In the transportation setting, this may lead to short and fast jerky vehicular motion, which could further degrade ride quality. Finally, note the effect of the change in $\gamma_2$: the change in amplitude is larger when $\gamma_2$ is increased from $0.5$ to $0.9,$ as opposed to when it is varied from $0$ (\emph{i.e.,} CCFM) to $0.5.$

\section{Concluding Remarks}
\label{section:Conclusions}
In this paper, we brought forth the effects of DAF on the qualitative dynamical properties of a platoon of vehicles traversing a straight road without passing. Motivated by the positive impact of DAF in several applications, we began by incorporating DAF in the evolution equations of the CCFM. We then derived the necessary and sufficient condition for local stability of the CCFM-DAF, and showed that DAF shrinks the locally stable region. Next, we proved that DAF preserves the \emph{manner} in which the CCFM loses local stability by proving the transversality condition of the Hopf spectrum. This results in the emergence of limit cycles in system dynamics. We also showed that DAF increases the frequency of emergent limit cycles.

We then made use of a suitable linear transformation to obtain insight into the local bifurcation properties of the CCFM-DAF by analyzing the analogous properties of the CCFM. Specifically, we proved that the CCFM-DAF undergoes a Hopf bifurcation if and only if the CCFM does. Additionally, this allowed us to study the type of Hopf bifurcation and the asymptotic orbital stability of the emergent limit cycles for the CCFM-DAF by analyzing the CCFM. We then studied the effect of DAF on three important properties; namely, non-oscillatory convergence, string stability and robust stability. Specifically, we proved that DAF destroys the non-oscillatory property of solutions of the CCFM, increases the risk of string instability and makes the CCFM less resilient to parametric uncertainty. In the process, we derived a sufficient condition for string stability and a necessary condition for robust stability of the CCFM-DAF. Finally, we showed that DAF decreases the amplitude of emergent limit cycles using a numerically-constructed bifurcation diagram. This served to show that DAF may induce fast and short jerks in the vehicular motion, which could degrade ride quality. Thus, our work reported a practically-relevant application wherein DAF degrades the performance across several measures of interest.

\section*{Acknowledgements}
This work is undertaken as a part of an Information Technology Research Academy (ITRA), Media Lab Asia, project titled ``De-congesting India's transportation networks.'' The authors are also thankful to Debayani Ghosh and Sreelakshmi Manjunath for many helpful discussions.


\begin{thebibliography}{99}








\bibitem{DC}
D. Chowdhury, L. Santen and A. Schadschneider, ``Statistical physics of vehicular traffic and some related systems,'' \emph{Physical Reports}, vol. 329, pp. 199-329, 2000.

\bibitem{DCG}
D.C. Gazis, R. Herman and R.W. Rothery, ``Nonlinear follow-the-leader models of traffic flow,'' \emph{Operations Research}, vol. 9, pp. 545-567, 1961.

\bibitem{DH}
D. Helbing, ``Traffic and related self-driven many-particle systems,'' \emph{Reviews of Modern Physics}, vol. 73, pp. 1067-1141, 2001.

\bibitem{EAU}
E.A. Unwin and L. Duckstein, ``Stability of reciprocal-spacing type car-following models,'' \emph{Transportation Science}, vol. 1, pp. 95-108, 1967.

\bibitem{GKK}
G.K. Kamath, K. Jagannathan and G. Raina, ``Car-following models with delayed feedback: local stability and Hopf bifurcation,'' in \emph{Proceedings of the $53^{rd}$ Annual Allerton Conference on Communication, Control and Computing}, 2015.

\bibitem{multi}
G.K. Kamath, K. Jagannathan and G. Raina, ``Impact of delayed acceleration feedback on the reduced classical car-following model,'' in \emph{Proceedings of the 2016 IEEE Conference on Control Applications (CCA)}, 2016.

\bibitem{arx}
G.K. Kamath, K. Jagannathan and G. Raina, ``Stability, convergence and Hopf bifurcation analyses of the classical car-following model,'' \emph{arXiv preprint arXiv:1607.08779}, 2016.

\bibitem{ifac}
G.K. Kamath, K. Jagannathan and G. Raina, ``String and robust stability of connected vehicle systems with delayed feedback,'' to appear in \emph{Proceedings of the $14^{th}$ IFAC Workshop on Time Delay Systems}, 2018.




\bibitem{GO}
G. Orosz and G. St$\acute{\text{e}}$p$\acute{\text{a}}$n, ``Subcritical Hopf bifurcations in a car-following model with reaction-time delay,'' \emph{Proceedings of the Royal Society A}, vol. 642, pp. 2643-2670, 2006.



\bibitem{HG}
H. Gomi and M. Kawato, ``Neural network control for a closed-loop system
using feedback-error-learning,'' \emph{Neural Networks}, vol. 6, pp. 933–946, 1993.


\bibitem{GL}
I. Gy$\ddot{\text{o}}$ri and G. Ladas, ``Oscillation Theory of Delay Differential Equations With Applications,'' \emph{Clarendon Press}, 1991.

\bibitem{JI}
J.I. Ge and G. Orosz, ``Dynamics of connected vehicle systems with delayed acceleration feedback,'' \emph{Transportation Research Part C}, vol. 46, pp. 46-64, 2014.

\bibitem{HL}
J.K. Hale and S.M.V. Lunel, ``Introduction to Functional Differential Equations,'' \emph{Springer-Verlag}, 2011.

\bibitem{LP}
L.E. Peppard, ``String stability of relative-motion PID vehicle control systems,'' \emph{IEEE Transactions on Automatic Control}, vol. 19, pp. 579-581, 1974.










\bibitem{MBD}
M. Bando, K. Hasebe, K. Nakanishi and A. Nakayama, ``Analysis of optimal velocity model with explicit delay,'' \emph{Physical Review E}, vol. 58, pp. 5429-5435, 1998.





\bibitem{RDD}
R.D. Driver, ``Ordinary and Delay Differential Equations,'' Springer-Verlag, 1977.

\bibitem{REC}
R.E. Chandler, R. Herman and E.W. Montroll, ``Traffic dynamics: studies in car following,'' \emph{Operations Research}, vol. 6, pp. 165-184, 1958.

\bibitem{REW}
R.E. Wilson and J.A. Ward, ``Car-following models: fifty years of linear stability analysis - a mathematical perspective,'' \emph{Transportation Planning and Technology}, vol. 34, pp. 3-18, 2011.

\bibitem{RH}
R. Herman, E.W. Montroll, R.B. Potts and R.W. Rothery, ``Traffic dynamics: analysis of stability in car following,'' \emph{Operations Research}, vol. 7, pp. 86-106, 1959.


\bibitem{ROS}
R. Olfati-Saber and R.M. Murray, ``Consensus problems in networks of agents with switching topology and time-delays,'' \emph{IEEE Transactions on Automatic Control}, vol. 49, 2004.


\bibitem{RS}
R. Sipahi and S.I. Niculescu, ``Analytical stability study of a deterministic car following model under multiple delay interactions,'' in \emph{Proceedings of Mechanical and Industrial Engineering Faculty Publications}, 2006.




\bibitem{TI}
T. Insperger, J. Milton and G. St$\acute{\text{e}}$p$\acute{\text{a}}$n, ``Acceleration feedback improves balancing against reflex delay,'' \emph{Journal of the Royal Society Interface}, Vol. 10, pp. 1-12, 2013.

\bibitem{TV}
T. Vyhl$\acute{\text{i}}$dal, N. Olgac and V. Ku\v{c}era, ``Delayed resonator with acceleration feedback -- Complete stability analysis by spectral methods and vibration absorber design,'' \emph{Journal of Sound and Vibration}, vol. 333, pp. 6781-6795, 2014.

\bibitem{XZ}
X. Zhang and D.F. Jarrett, ``Stability analysis of the classical car-following model,'' \emph{Transportation Research Part B}, vol. 31, pp. 441-462, 1997.




















%

\end{thebibliography}
\end{document}